\documentclass[twocolumn,prb]{revtex4}
\usepackage{amssymb}
\usepackage{amsmath}
\usepackage{epsfig}

\def\bra#1{\mathinner{\langle{#1}|}}
\def\ket#1{\mathinner{|{#1}\rangle}}

\def\leer{\varnothing}

\def\id3{ {\mathbf{1}_{3\times 3}}}
\def\ln {\text{ln}}

\begin{document}

\title{DMRG studies of the effect of constraint release on the viscosity of polymer melts}
\author{Matthias Pae\ss ens}
\email{m.paessens@fz-juelich.de}
\author{Gunter M. Sch\"utz}

\affiliation{Institut f\"ur Festk\"orperforschung, Forschungszentrum J\"ulich - 
52425 J\"ulich, Germany}
\date{\today}

\begin{abstract}
The scaling of the viscosity of polymer melts is investigated with regard to 
the molecular weight. 
We present a generalization of the Rubinstein--Duke model, which takes constraint 
releases into account and calculate the effects on the viscosity by the use of the Density
Matrix Renormalization Group (DMRG) algorithm. Using input from Rouse theory the rates for 
the constraint releases are determined in a self consistent way. We conclude that shape 
fluctuations of the tube caused by constraint release are not a likely candidate for improving
Doi's crossover theory for the scaling of the polymer viscosity.

\end{abstract}

\maketitle

\section{Introduction}
The viscosity of polymer melts has been investigated intensively.\cite{ferry}
The models for describing this behavior, however, are not yet satisfying. Experiments show that the 
viscosity $\eta$ scales like $M^{3.3\pm 0.1}$, where $M$ is the molecular weight. This behavior is valid
for several decades of the molecular weight. An early approach was the reptation model by de Gennes 
\cite{degennes} which yields $\eta \propto M^3$ in the limit of infinitely long polymers. For short polymers
Doi\cite{doiedwards} calculated the viscosity by taking tube length fluctuations into account
 and found a region
of $M$ where the scaling is of the correct size -- but this region appears to be much too small. 
A discrete model
for reptation which includes tube length fluctuations is the Rubinstein--Duke (RD) model.\cite{rubinstein,duke}
Recently it was shown \cite{carlon} that this model does not only provide a 
region where a scaling of $M^{3.3\pm 0.1}$ can be found, as already shown by Rubinstein, \cite{rubinstein} 
but also that this model shows a crossover to the reptation exponent 3. But again the 
mass region where
the correct exponent is valid does not exceed one order of magnitude. 
In this paper we extend the RD model in order to investigate whether  
constraint release (CR) broadens that region. While good models for the
non--linear regime of viscosity can be constructed by using CR, \cite{hua,likhtman} there was only
little success in the linear regime.\cite{rotstein}\\

\section{theory}
\subsection{RD model}
\begin{figure}
{   \begin{center}
\resizebox*{!}{3.8cm}{\includegraphics{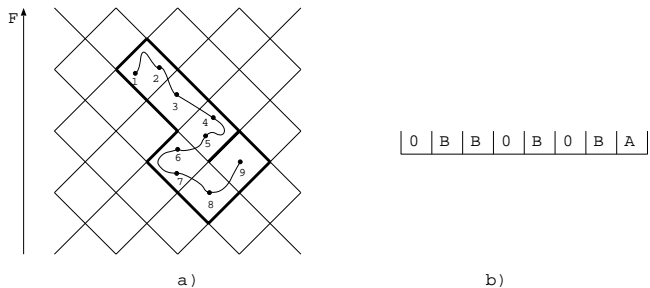}}
    \end{center} }
\caption{a) The repton model in two dimensions: the circles represent the reptons; the
primitive path is marked by the bold lines. b) Projection onto one dimension.}
\label{reptonmodel}
\end{figure}

In order to investigate the role of tube length fluctuations in reptation theory Rubinstein introduced a
discrete ``repton'' model which allows the description of three--dimensional reptation dynamics 
by a one--dimensional lattice gas.\cite{rubinstein}
Duke generalized the model to the case that an external electric field acts on a charged polymer\cite{duke}
which allows a good description of the diffusion constant in gel electrophorese 
experiments.\cite{euroschuetz} However, also in the absence of a field the generalization by Duke
is useful in so far as it provides a reference axis along which the displacement of the polymer chain
as a whole can be monitored. This makes it possible to calculate the diffusion coefficient and
the viscosity from the model, without resorting to independent hypotheses.\\
Rubinstein assumes that the constraints of the other polymers divide space into cells which form
a $d$--dimensional regular cubic lattice.
The polymer occupies a series of adjacent cells, the ``primitive path''.
It is not possible for the polymer to traverse the edges of the cells (in two dimensions: the lattice
points) so that only the ends of the polymer can enter new cells. The polymer is divided into segments whose
length is of the order of the lattice constant, the number of segments is proportional to the length of 
the polymer or the molecular weight. The orientation of the lattice is introduced by Duke in a 
way that the electric
field is diagonal to the lattice, i.e. in three dimensions in the (111)--direction.\\
A segment -- called a ``repton'' -- is allowed to jump into an adjacent cell according to the 
following rules: \\
\begin{enumerate}
\item The reptons in the bulk are only allowed to jump along the primitive path.
\item No cell in the interior of the primitive path may be left empty.
\item The ends move freely insofar rule 2 is respected. If an end repton occupies the cell
alone, it can only retract in the cell of the adjacent repton. If the adjacent repton is in
the same cell, the end repton may enter any of the $2d$ surrounding cells.
Reptons in the bulk jump with the same probability as 
reptons at the end into occupied cells.
\end{enumerate}
Rule 1 ensures that the polymer does not traverse the edges of the cells. Rule 2 is motivated by the
fact that the segments are of the size of the lattice constant. Finally rule 3 reflects the fact that
there are more free adjacent cells for an end repton than occupied ones. If one considers a field
$F\neq 0$ the ratio of the probabilities to jump in respectively against the direction of the field 
is proportional to the Boltzmann weight. In the following we will only focus on the case without field,
we only need the reference axis.\\
As the shape of the primitive path is not affected by the movement of the polymer and only the ends are
created or annihilated this model can be mapped onto one dimension. 
To this end a particle is assigned to each bond between reptons; 
starting on one end, a bond into the direction of the field is identified with an $A$--particle,
against the field with a $B$--particle and finally a bond without change of potential, i.e. a bond between
two reptons in one cell, by a vacancy $\varnothing$ (Fig. \ref{reptonmodel}).\footnote{In the Rubinstein
model only two cases are distinguished; a bond between two cells is identified by a particle and a bond 
within a cell is identified by a vacancy.}
\\
These particles are residing on a chain whose number of sites $N$ is the number of bonds, thus the
number of reptons minus one. As for each site three states are possible this model can be
identified with a ``quantum'' spin--one chain.\cite{schuetz,barkema} We define 
\begin{equation}
A\equiv \begin{pmatrix}1\\0\\0\\ \end{pmatrix},\leer\equiv \begin{pmatrix}0\\1\\0\\ \end{pmatrix},
 B\equiv \begin{pmatrix}0\\0\\1\\ \end{pmatrix}.\nonumber 
\end{equation}
We label this set of states with $X=\{A,\leer,B\}$. A state of a chain of length $N$ can be
considered as an element $\ket\eta$ of the tensor base ${\mathbf X}=X^{\otimes N}$; $\ket\eta$
is constructed by the tensor product of the three component vectors for the individual sites.\\
The dynamics of this one dimensional model is:
\begin{enumerate}
\item $A$--particles may exchange with vacancies $\leer$:  $A\leer \leftrightharpoons \leer A$.
\item $B$--particles may exchange with vacancies $\leer$:  $B\leer \leftrightharpoons \leer B$. 
\item $A$--particles may {\em not} exchange with $B$--particles.
\end{enumerate}
By these rules it is guaranteed that the shape of the primitive path is conserved. The next rules
define the boundary dynamics:
\begin{enumerate}
\addtocounter{enumi}{3}
\item \label{itA0} At the ends of the chain $A$--particles may be annihilated:\\ $A\rightarrow \leer$.
\item \label{itB0} At the ends of the chain $B$--particles may be annihilated:\\ $B\rightarrow \leer$.
\item \label{it0A} At the ends of the chain $A$--particles may be created: $\leer \rightarrow A$.
\item \label{it0B} At the ends of the chain $B$--particles may be created: $\leer \rightarrow B$.
\end{enumerate}
The processes 1, 2, 4 and 5 take place with the same probability; the processes
6 and 7 are $d$--times more probable. \\
In this way a stochastic interacting particle system on a one dimensional chain 
has been defined. As the transitions
are independent of the previous history this process is Markovian.\\

\subsection{Quantum Hamiltonian}
A convenient way to describe the process mathematically is the quantum Hamiltonian formalism which we
will present here shortly, for details see Ref. \onlinecite{schuetz}.\\
The probability to be in state $\ket\eta$ at time $t$ is labeled by $P_\eta(t)$. These probabilities
of the individual states can be combined to a vector: $\ket{P(t)}=\sum P_\eta(t) \ket\eta$.
Due to the conservation of probability the entries of the vector $\ket{P(t)}$ sum up to 1 at
any time $t$. With this definition the master equation can be written as: 
\begin{equation}
\frac{d}{dt}\ket{P(t)}=-H\ket{P(t)},
\end{equation}
with the stochastic generator 
\begin{equation}
H= -\sum_\eta\sum_{\eta^\prime\neq\eta} w_{\eta^\prime \rightarrow \eta} \ket\eta\bra{\eta^\prime}
    + \sum_\eta\sum_{\eta^\prime\neq\eta}w_{\eta \rightarrow \eta^\prime} \ket\eta\bra{\eta}.
\end{equation}
Here $w_{\eta^\prime \rightarrow \eta}$ is the transition probability from state $\ket{\eta^\prime}$
to $\ket{\eta}$. In other words the off diagonal elements of the matrix $H$ are the negative transition
rates between the respective states and the diagonal elements are the sum of the rates
leading away from the respective state. \\
The creation operators $a^\dagger$ and $b^\dagger$ are defined by $a^\dagger \leer =A$ and 
$b^\dagger \leer =B$, acting on any other state yields zero. 
The annihilation operators $a$ and $b$ are defined by $aA=\leer$ and $bB=\leer$, again
acting on any other state yields again zero. Finally we define the number operators $n^A=a^\dagger a$,
$n^B=b^\dagger b$ and $n^\leer=1-n^A-n^B$. By these definitions the stochastic generator $H$ of
the RD model reads: \cite{barkema}
\begin{equation}
H=b_1(d)+b_N(d)+\sum_{n=1}^{N-1}u_n
\label{H_rubinstein_duke}
\end{equation}
with
\begin{eqnarray*}
b_n(d)&=&d\left[n_n^\leer -a_n^\dagger+n_n^\leer -b_n^\dagger\right]+n_n^A-a_n+n_n^B-b_n \\
u_n&=&n_n^An_{n+1}^\leer -a_na_{n+1}^\dagger  +   n_n^Bn_{n+1}^\leer -b_nb_{n+1}^\dagger \\
  &+& n_n^\leer n_{n+1}^A-a^\dagger_na_{n+1}  +   n_n^\leer n_{n+1}^B-b^\dagger_nb_{n+1}.
\end{eqnarray*}
Again, $d$ labels the lattice dimension. We have chosen the time scale such that the hopping
rate in the bulk equals unity.

\subsection{Calculation of the viscosity}
The viscosity is proportional to the longest
relaxation time of the stochastic generator $H$ which is the inverse energy gap.\cite{doiedwards}  
Due to the conservation of probability the ground state of a stochastic generator
has the eigenvalue zero, so that we only need to calculate the first eigenvalue. \\
A very efficient algorithm to calculate the lowest excitations of quantum spin chains
is the density matrix renormalization group (DMRG) algorithm.\cite{white,noack} For the standard RD model
the usefulness of this algorithm has been demonstrated,\cite{carlon} so that
we employ this algorithm as well for the modifications of the RD model, to be introduced in what follows.\\
The Hilbert space of the spin chain grows exponentially with the number sites $N$; the number of states
is $3^N$, so that for $N=50$ --- the maximum number of sites considered in our calculations ---
the number of states is of the order $10^{23}$. 
By this it is reasonable to project all quantities onto a subspace consisting of the most important states. 
The difficulty is to find out which states are the ``most important''. The DMRG algorithm is a method
which provides a choice of states which is optimal in terms of a maximum of the probability with which
the states contribute to the target state.\cite{white} 
This probability is gained by diagonalizing the density
matrix. 
As the diagonalization of the density matrix for the whole system would be as laborious as the 
direct diagonalization of the hamiltonian the system is build up stepwise. Starting with a small
system (e.g. 2 states) one adds iteratively two states until the system has reached the searched
size. In each step the Hilbert space is reduced to the subspace of the most probable states, so 
that the system size increases while the dimension of the matrices remain constant.

\subsection{Constraint release}
\begin{figure}
{   \begin{center}
\resizebox*{!}{5cm}{\includegraphics{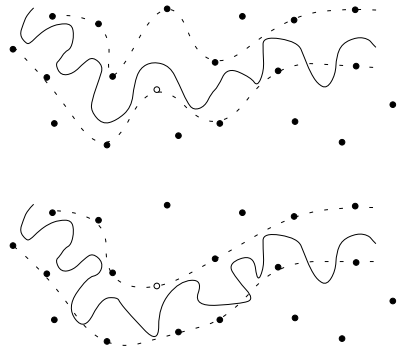}}
    \end{center} }
\caption{A constraining polymer vacates for a short time its site and thus makes 
a constraint release event possible.}
\label{CR1}
\end{figure}

\begin{figure}
{   \begin{center}
\resizebox*{!}{4cm}{\includegraphics{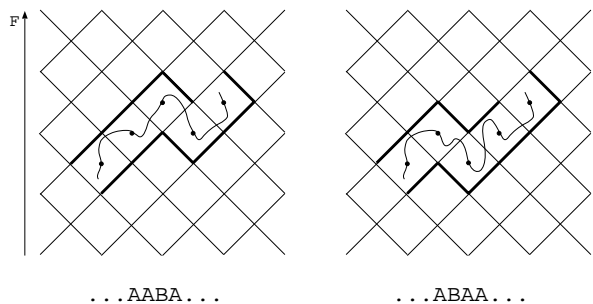}}
    \end{center} }
\caption{Constraint release by $AB$--permutation in the RD model.}
\label{CR2}
\end{figure}

The RD model as well as standard reptation theory is based on the assumption 
that the polymer moves in a fixed network
formed by the surrounding polymers. This assumption is at best justifiable for a single polymer immersed
in a gel. But for polymer melts, for which the viscosity is measured, it should be taken into account that
the surrounding polymers reptate themselves. In terms of the Rubinstein model this means that the
lattice itself is subjected to fluctuations. \\
We consider the following model of lattice fluctuations: Imagine that a constraining polymer moves so far that
the constraint for the investigated polymer is released so that it can move freely in this region.
After a short time the constraining polymer returns or an other polymer has taken its place so that
the free movement in this region is again prevented. The lattice has regained its originally
structure but the primitive path may have changed in the bulk (Fig. \ref{CR1}). 
This is what we call a  ``constraint release'' (CR) event. 
\\
A straightforward
implementation of this mechanism is shown in Fig. \ref{CR2}. The turning of bend of the
primitive path equals to the permutation of an $AB$--pair. At the ends of the path 
$A$--particles can be transformed into $B$--particles and vice versa which corresponds 
to a CR event at the ends. 
\\
To construct the extended Hamiltonian which takes these mechanisms into account we introduce
an operator which transforms $B$--particles into $A$--particles:
\begin{equation}
c=a^\dagger b; \quad c^\dagger=b^\dagger a,
\end{equation}
the adjoint operator effects the reversed process. The corresponding diagonal elements are
build up by $n^B$ respectively $n^A$. Now the new Hamiltonian can be written as:
\begin{equation}
\begin{split}
H^{CR}&=H+g_1(\alpha)+g_N(\alpha)+\sum_{n=1}^{N-1} v_n(\alpha) \\
g_n(\alpha)&= \alpha \left[ n_n^B-c_n + n_n^A-c_n^\dagger \right]\\
v_n(\alpha)&= \alpha \left[ n_n^B n_{n+1}^A-c_n c^\dagger_{n+1} + n_n^A n_{n+1}^B - c_n^\dagger c_{n+1}\right],
\end{split}
\end{equation}
with the Hamiltonian $H$ from Eq. (\ref{H_rubinstein_duke}), and $\alpha$ labels the rate for the
CR process.

\section{Calculations and results}

\begin{figure}{
\begin{center}
\epsfig{figure=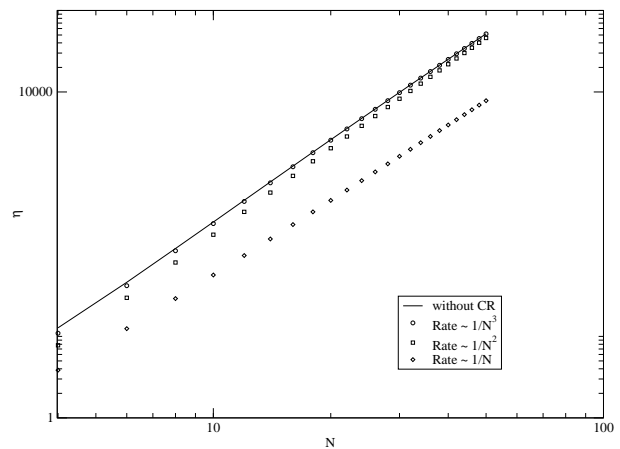,scale=0.3,angle=-90} 
\caption{The viscosity ($\eta \propto E_N^{-1}$) for several CR rates of $AB$ permutation. The solid line
shows the result for the standard RD model without CR.}\label{eta_cr}
\end{center}
 }
\end{figure}

\begin{figure}{
\begin{center}
\epsfig{figure=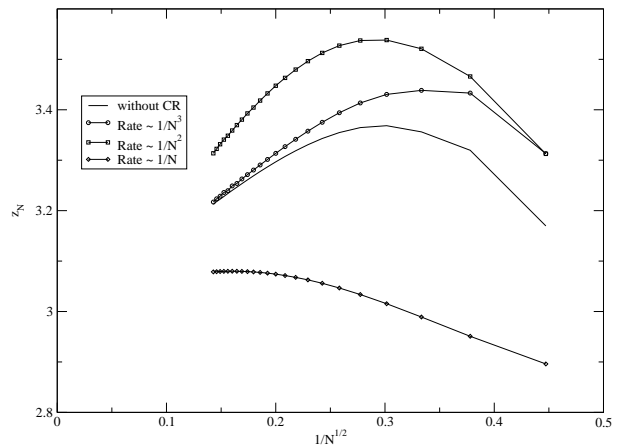,scale=0.3,angle=-90} 
\caption{The effective exponent $z_N$ for several CR rates of $AB$ permutation. The solid line
shows the result for the standard RD model without CR.}\label{exp_cr}
\end{center}
 }
\end{figure}

\begin{figure}{
\begin{center}
\epsfig{figure=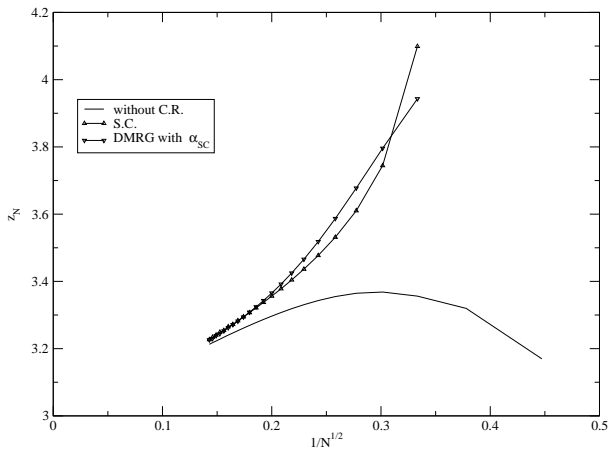,scale=0.3,angle=-90} 
\caption{The effective exponent $z_N$ of the viscosity with self--consistent CR.}\label{exp_sc}
\end{center}
 }
\end{figure}

In order to investigate the dependence of the viscosity on the CR rate $\alpha$ we 
performed DMRG calculations for the rates $1/N$, $1/N^2$ and $1/N^3$. The choice of
these rates is based on two estimates. The first is very simple: A constraint is
released when an end repton retracts back into the tube. The probability that the
constraint is exerted by an end repton is $2/N$, provided that the probability for
the presence of a segment is equally distributed. The hopping of a single repton is
a process of rate one and so we find that the rate for a CR is of
the order $1/N$. For the second estimate we assume that a CR is caused by a 
tube renewal. Later in this paper we will show the calculation of this estimate within
the Rouse model which leads to the result that the rate $\alpha$ is proportional to the
inverse relaxation time and scales therefore with $1/N^3$. \\
The viscosity which was calculated with these rates for the mechanism of
AB exchange is plotted in Fig. \ref{eta_cr}. It can
bee seen that the viscosity decreases with increasing CR rate which is reasonable because
the relaxation of the tube is accelerated by this process. For the investigation of the
scaling it is more useful to plot the local slope or the effective exponent
\begin{equation}
z_N=\frac{\ln\, \tau_{N+1} - \ln\, \tau_{N-1}}{\ln (N+1) - \ln (N-1)}, 
\label{eq_zn}\end{equation}
against $1/\sqrt{N}$ as introduced in Ref. \onlinecite{carlon}. The choice of the abscissa is motivated
by the formula of Doi\cite{doiedwards} which predicts a correction to the $N^3$--scaling in the 
order of $1/\sqrt{N}$ by taking tube length fluctuations into account. It should be mentioned that in this paper
$N$ labels the number of segments which is proportional to the length of the polymer while
in Ref. \onlinecite{carlon} it labels the number of reptons which is the number of segments plus one.
Surprisingly the latter interpretation shows a better agreement with the Doi formula while the former shows
a better agreement with the experiments since the range where $z_N \approx 3.3\pm 0.1$ is much
broader.
 \\
The influence on the effective exponent $z_N$ is plotted in Fig. \ref{exp_cr}. The influence is
non--monotonic in $\alpha$: While the rate $1/N$ causes an obvious shift down in comparison with the data
without CR, the data of rate $1/N^2$ is located above the curve without CR. The data of 
rate $1/N^3$ is also located above the curve without CR but at a smaller distance. \\
We remark that from a theoretical point of view also the rate $\alpha=1$ is interesting. 
If $A$-- and $B$--particles
exchange with the same rate as particles with vacancies, the particle can diffuse freely, as they
feel no longer restrictions. This means that the reptation model transforms into the Rouse model
as the tube can change its form freely. This is why rate $\alpha=1$ represents a possibility to
verify the model: The relaxation time should scale in the limit $N \to \infty$ as $N^2$. But the
DMRG calculation of the effective exponent with this rate yields a diverging curve. On the other
hand, if the
second excitation is considered, one can see that it scales as $N^{-2}$. This suggests
that the first excitation is caused by an other relaxation mechanism and the
second excitation yields the desired time, which scales with the expected exponent whereby
the transition to Rouse dynamics is verified. 
 \\
The bounds $1/N > \alpha > 1/N^3$ are rather weak and a self--consistent approach is needed for 
determining the appropriate scaling of the rate $\alpha$. 
As a CR event is caused by the tube renewal of a polymer,
this process is not independent of the relaxation time. The relaxation time itself is affected by 
the CR rate. So the rate is physical, if the relaxation time, which is calculated using this rate, 
yields the same CR rate. 
\begin{figure}{
\begin{center}
\epsfig{figure=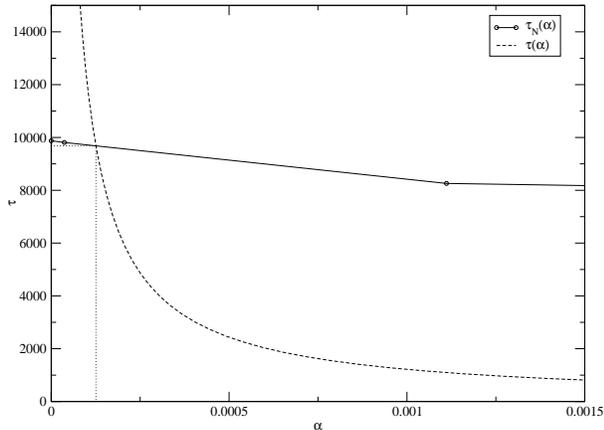,scale=0.3,angle=-90} 
\caption{Relaxation time versus CR rate $\alpha$; $\alpha_{SC}$ is given by the intersection
point of the curves $\tau_N$ and $\tau$ (here exemplified for $N=30$).}\label{fig_tau_alpha}
\end{center}
 }
\end{figure} \\
To calculate the self consistent rate $\alpha_{SC}$ we proceed as follows: The dependence of
the CR rate on the relaxation time $\alpha(\tau)$ is estimated by an analytical calculation using Rouse
dynamics.
The DMRG calculation with the rates mentioned above yields directly the dependence
of the relaxation time on the CR rate $\tau_N(\alpha)$ for the respective chain lengths. The
inverse function of $\alpha(\tau)$, $\tau(\alpha)$, is compared with $\tau_N(\alpha)$: The intersection
 of the curves yields the self consistent rate $\alpha_{SC}$ in first approximation.\\
First we calculate the dependence of the CR rate on the relaxation time. The polymer 
whose relaxation time we want to calculate finally is hindered in his free movement by an other
polymer. The position which exerts the constraint may have the distance $s$ from one end of the 
constraining polymer. So, the constraint is released when the polymer moves either the distance
$s$ into the one direction or the distance $N-s$ into the other direction, where $N$ labels
the length of the polymer. In Ref. \onlinecite{doiedwards} an expression is presented which
provides the distribution $\psi(s,t)$ of the probability that the tube segment $s$ was at 
time $t$ not yet reached by the ends:
\begin{equation}
\psi(s,t)=\sum_{p,odd}\frac{4}{p\pi}\sin\left(\frac{p\pi s}{N}\right)e^{-p^2 t/\tau},
\end{equation}  
with the relaxation time $\tau$. The calculation of this formula bases on the assumption that the
polymer diffuses in between the tube due to Rouse dynamics. The complementary probability distribution
$\phi(s,t)$, which indicates the probability that at time $t$ the segment $s$ is already reached by
one of the ends is then
\begin{equation}
\phi(s,t)=1-\psi(s,t).
\end{equation}
The derivate with respect to time 
\begin{equation}
f(s,t)=\frac{\partial \phi}{\partial t}
\end{equation}
is the first passage time density, i.e. the probability per time that the segment $s$ is reached
by one end just at time $t$. Now we can specify the mean first passage time
\begin{equation}
\bar\mu(s)=\int_0^\infty dt\,t\,f(s,t)
\end{equation}
which indicates how long it takes in average until segment $s$ is reached by an end. Finally
we average over all $s$ provided that each segment builds up an entanglement with the same probability
\begin{equation}
\mu=\frac{1}{N}\int_0^N ds\,\bar\mu(s).
\label{mu}
\end{equation}
This is the desired time: In average the release of a constraint will take the time $\mu$,
the CR process takes place with rate $\alpha=\mu^{-1}$. The calculation of the integrals
is elementary mathematics and we only show the result:
\begin{equation}
\mu=0.822 \tau \Rightarrow \alpha = 1.22 \tau^{-1}.
\end{equation}
This calculation is based on a continuous description of the polymer. 
However, we are considering
a lattice model of reptation so that boundary effects might not be taken into account correctly 
by the continuum expression (\ref{mu}).
Therefore we performed the above calculation as well for the discrete case. As the result for
the discrete case converges quickly to the continuous case but consists of only
numerical evaluable series, we restricted ourselves to the continuous case here.
There is an error of less than 2 per cent for as little as 10 reptons.
\begin{figure}{
\begin{center}
\epsfig{figure=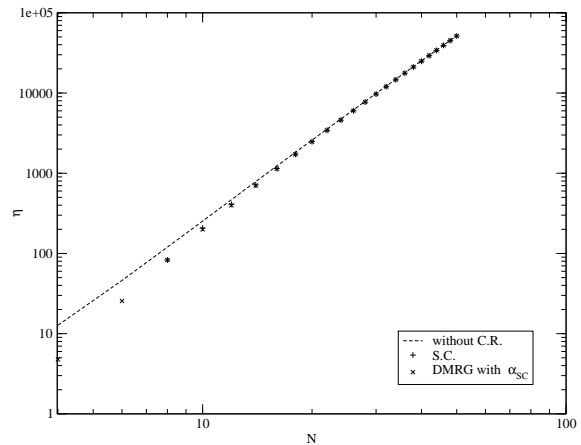,scale=0.3,angle=-90} 
\caption{The new DMRG calculation of viscosity -- a good coincidence with the data of the intersection points can be seen.}\label{eta_sc}
\end{center}
 }
\end{figure}\\
Fig. \ref{fig_tau_alpha} shows the curve $\tau(\alpha)=1.22\alpha^{-1}$ and the relaxation times
for the CR rates $\alpha=1/N^3$ and $1/N^2$ exemplified for $N=30$, the data point for $\alpha=1/N$ 
is out of the plotted range. 
In order to investigate the dependence of the relaxation time on the CR rate we interpolated 
the data points linearly by fitting them with splines. The data points are is well 
approximated by this interpolation in
the region where the intersection point is expected. In this way we determined the self consistent
relaxation times and rates for the lengths $N=8\ldots 50$. As the lengths $N=4$ and $N=6$ do not
lead to reasonable intersection points we did not take them into account. To verify the self 
consistence a new DMRG run was done with a fit of the determined rates; as can be seen
in Fig. \ref{eta_sc} the resulting relaxation times coincide well with the ones determined
by the self consistence condition, so that the linear interpolation represents a sufficiently
good approximation. \\
In Fig. \ref{exp_sc} the effective exponent is shown for the self consistent relaxation times and for
the DMRG calculation. The exponent is shifted to higher values in the region of small $N$, but a
broadening of the region where $z_N\approx 3.3$ could not be observed.\\
   \\

The Rubinstein model is not microscopic and hence there is some freedom in implementing microscopic
process such as CR. A different mechanism\cite{euroschuetz} leads to the creation and annihilation
of particles in the bulk. We performed similar analysis for this mechanism with qualitatively similar
results.\cite{diploma}\\
We conclude that Rouse--based calculations do not lead to a proper description of crossover behavior 
in the viscosity of polymer melts when Rouse theory is used as a self consistency input in the 
mesoscopic and generally quite successful Rubinstein model for reptation. We cannot rule
out that a fully self--consistent implementation of CR (without Rouse assumption) leads to a
crossover regime closer to empirical evidence, but the broad range of CR rates studied here
suggests that bulk shape fluctuations of the tube which result from constraint release do not
significantly broaden the crossover range of the viscosity.

\end{document}